\documentclass{jfm}

\usepackage{graphicx}
\usepackage{epstopdf, epsfig}
\usepackage{amsmath}
\usepackage{upgreek}
\usepackage[dvipsnames]{xcolor}

\usepackage{hyperref}
\hypersetup{colorlinks=true,linkcolor=blue,urlcolor=blue,citecolor=blue}

\newcommand{\Ma}{\mbox{\textit{Ma}}}  
\newcommand{\Bq}{\mbox{\textit{Bq}}}  
\newcommand{\Pe}{\mbox{\textit{Pe}}}  

\title{Shear flow over a surface containing a groove covered by an incompressible surfactant phase}
\author{Tobias Baier\aff{1}\corresp{\email{baier@nmf.tu-darmstadt.de}}, Steffen Hardt\aff{1}}

\affiliation{\aff{1} Fachbereich Maschinenbau, Technische Universit\"at Darmstadt, 64287 Darmstadt, Germany}

\shorttitle{Shear flow over a groove covered by incompressible surfactant}
\shortauthor{T. Baier and S. Hardt}

\begin{document}
	
\maketitle
	
\begin{abstract}
We study shear-driven liquid flow over a planar surface with an embedded gas-filled groove, with the gas-liquid interface protruding slightly above or below the planar surface. The flow direction is along the groove, taken to be much longer than wide, and the gas-liquid interface is assumed to be covered by an incompressible surface fluid, representing a surfactant phase. Using the incompressiblity condition for the surface fluid, the equations of motion and corresponding boundary conditions for the liquid phase are obtained by minimizing the dissipation rate. Assuming a moderate deformation of the interface, a domain perturbation technique with the maximal deformation as the small parameter is employed. The Stokes equation in the liquid phase under corresponding boundary conditions is solved to second order in the deformation using the Keldysh-Sedov formalism. The obtained analytical results are compared with numerical calculations of the same problem, allowing an assessment of the limits of validity of the expansion. While on a planar gas-liquid interface no flow is induced, a recirculating flow is observed on an interface protruding slightly above or below the planar surface. The study sheds light onto the mobility of curved gas-liquid interfaces in the presence of surfactants acting as an incompressible surface fluid.
\end{abstract}


\section{Introduction}\label{sec:intro}

The presence of surfactants can significantly impact the flow in the vicinity of fluid interfaces. For bounded fluid interfaces, the convection of interface--bound surfactant molecules can lead to their stacking up at stagnation points, which in turn results in Marangoni stress, effectively rendering the interface incompressible through stress directed along the negative concentration gradient \citep{Manikantan_2020}. Well known examples are the formation of a stagnant surfactant film in front of an obstacle piercing the surface of a flowing liquid \citep{Merson_1965, Scott_1982, Harper_1992} or the reduced mobility of drops or bubbles moving through a liquid due to an increased surfactant concentration on the downstream portion of their interface \citep{Savic_1953, Levich_1962, Davis_1966, Sadhal_1983}.

Special attention has been given to liquid flow over 'superhydrophobic' surfaces with gas-filled (or liquid-filled) cavities due to their potential for drag reduction \citep{Rothstein_2010, Schonecker_2014, Lee_2016}. The influence of surfactants on such flows have been observed experimentally \citep{Kim_2012, Bolognesi_2014, Schaffel_2016, Peaudecerf_2017, Song_2018, Li_2020} and analyzed theoretically \citep{Gaddam_2018, Landel_2020, Baier_2021, Temprano_2021}. Also here, surfactant molecules stack up at stagnation points and can severely impede the drag reduction compared to expectations based on a surfactant-free situation. In the extreme case of large surfactant concentrations or low enough shear rates, Marangoni stress at the interface can completely balance the viscous shear stress, rendering the interface immobile \citep{Peaudecerf_2017, Baier_2021}. One such situation where this can occur is homogeneous shear flow over a flat gas-liquid interface embedded in an entirely flat surface, resulting in a constant shear rate on the interface. However, when the interface experiences an inhomogeneous shear stress, for example on a curved gas-liquid interface protruding into the channel \citep{Song_2018} or side-walls imposing an inhomogenous shear-field \citep{Li_2020}, recirculation zones can occur on the interface at sufficiently high surfactant concentrations. A similar phenomenon is observed when vesicles attached to a flat interface are subjected to shear flow, where recirculation becomes observable in the lipid bilayer enclosing the vesicle \citep{Woodhouse_2012, Honerkamp_2013}. In reality, flat gas-liquid interfaces are more an exception than a rule. Gas dissolution in the liquid or the pressure drop in a channel with superhydrophobic walls deform gas-liquid interfaces. It was shown experimentally that the interface curvature influences the flow over bubble mattresses very significantly \citep{Steinberger_2007, Karatay_2013}. By contrast, how the curvature of a surfactant-laden interface influences the flow along the interface has remained largely unexplored. 

We aim here at a qualitative understanding of shear driven flow along a long narrow gas-filled groove in a planar surface, similar to the experimental set-up employed by \cite{Song_2018}, by modeling the surfactants as an incompressible non-viscous surface-fluid. We thus assume that even small variations in surface concentrations impose such large Marangoni stresses on the bulk fluid that the surface flow is rendered virtually incompressible (large Marangoni number), while viscous stresses within the surface fluid are insignificant compared to viscous stresses in the bulk (small Boussinesq number) \citep{Manikantan_2020,Elfring_2016,Barentin_2000}. This simplification allows for an analytical solution of the flow field for small deflections of the gas-liquid interface away from the planar surface. For a shear-free gas-liquid interface without surfactants, the analogous situation of pressure-driven or shear-induced flow over surfaces with grooves containing gas-pockets with curved menisci has been analyzed numerically \citep{Ng_2011, Li_2017, Ageev_2018, Alinovi_2018}, analytically \citep{Sbragaglia_2007, Crowdy_2010, Crowdy_2015, Crowdy_2016, Kirk_2018, Asmolov_2018} and experimentally \citep{Karatay_2013, Kim_2019}. Similarly, thermocapillary flow along such surfaces due to a thermal gradient along the grooves \citep{Baier_2010} has received considerable attention in the case of curved menisci \citep{Kirk_2020, Yariv_2020, Yariv_2021}. Both of these situations bear some similarities to the one involving an incompressible surface-fluid, and similar techniques can be employed in their solution.

\section{Mathematical model}\label{sec:model}

\begin{figure}
	\centering
	\includegraphics[width=0.85\textwidth]{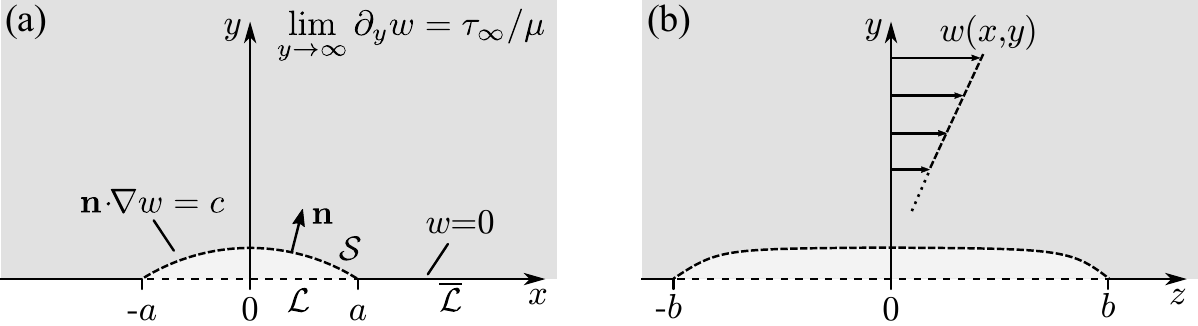}
	\caption{\label{fig:sketch} (a) Sketch of the geometry with a liquid above a surface with a long, narrow gas-filled cavity of width $2a$ and length $2b$. The gas-liquid interface, ${\cal S}$, is assumed to have the shape of a circular arc and laden with an insoluble surfactant. Application of a shear stress $\tau_\infty$ far away from the surface and in a direction normal to the $x$--$y$-plane drives a Couette flow along the groove. (b) Side view of the geometry. The region of interest is sufficiently far away from the ends of the groove, such that the deflection $h(x)$ of the gas-liquid interface can be considered independent of $z$.}
\end{figure}

The configuration under investigation is sketched in figure \ref{fig:sketch}, showing the view along the flow direction in (a) and a side-view in (b). A Newtonian liquid of viscosity $\mu$ is driven by application of a shear-stress $\tau_\infty$ along the $z$-axis far above a planar surface containing a single long, narrow gas-filled groove of width $2a$ and length $2b$, with $a \ll b $, oriented in the same direction as the applied shear-stress. The region of interest is an $x$--$y$-plane perpendicular to the flow direction not too close to the ends of the groove at $|z| = b$, such that the flow is described sufficiently well by a unidirectional velocity field $w(x,y)$ along the groove, unaffected by its finite length. In this region, the deflection $y=h(x)$ of the gas-liquid interface ${\cal S}$ can be assumed to have the form of a circular arc, independent of the $z$-coordinate along the groove, and we assume the flow to be slow enough that the shape of the gas-liquid interface is unaffected by viscous stresses acting on it. Furthermore, the viscosity of the gas in the cavity is assumed to be small enough that viscous stresses acting on the interface from the gas are negligible compared to stresses exerted by the liquid.

The gas-liquid interface is assumed to be covered by an incompressible, inviscid surface fluid, insoluble in the liquid. Since the flow is unidirectional, the mass conservation equation for the surface fluid becomes
\begin{equation}\label{eq:IntC_arc}
\int_{\cal S} w(x,y) ds = 0,
\end{equation}
where the integral is along the part ${\cal S}$ of the gas-liquid interface cut by the $x$--$y$-plane containing the liquid domain $\Omega$ of interest. In the limit of vanishing Reynolds number, the local momentum conservation for the liquid together with the local form of the mass conservation equation for the surface fluid can be obtained by minimizing the dissipation rate \citep{Batchelor_2000, Kim_2005} subject to the constraint \eqref{eq:IntC_arc}, 
\begin{equation}
\Upphi[w] = \mu \int_\Omega \boldsymbol{\nabla} w \cdot \boldsymbol{\nabla} w\, d^2 x + 2 \mu c \int_{\cal S} w ds
\end{equation}
where $c$ is a Lagrange multiplier. Expanding $\delta \Upphi = \Upphi[w+\delta w] - \Upphi[w]$ to first order in the variation $\delta w$, which is assumed to vanish on all boundaries of $\Omega$ except for ${\cal S}$, and setting it equal to zero, leads to
\begin{equation}\label{eq:Laplace}
\boldsymbol{\nabla}^2 w = 0,\quad \text{in }\Omega
\end{equation}
and, with the normal vector $\mathbf{n}$ pointing into the liquid domain,
\begin{equation}\label{eq:BC_arc}
\mathbf{n}\cdot\boldsymbol{\nabla} w = c, \quad \text{on } {\cal S}.
\end{equation}
The former of these is the Stokes equation for unidirectional shear driven flow, while the latter describes a constant shear-stress at the interface, reminiscent of a situation with thermal Marangoni flow along a groove with a constant temperature gradient along its surface \citep{Baier_2010}. However, while Marangoni flow is induced towards regions of higher surface tension only, here the boundary condition \eqref{eq:IntC_arc} requires a back-flow on parts of the interface. This is particularly striking when the interface is flat: in this case the shear rate dictated by the far-field condition,
\begin{equation}\label{eq:farField}
\partial_y w \to \tau_\infty/\mu, \quad \text{for } y\to\infty,
\end{equation}
extends all the way to the planar surface, since the velocity field and Lagrange multiplier
\begin{equation}
w_0(x,y)=\frac{\tau_\infty}{\mu} y, \qquad c_0=\frac{\tau_\infty}{\mu}
\end{equation}
solve the Laplace equation \eqref{eq:Laplace} with boundary conditions \eqref{eq:BC_arc}, \eqref{eq:farField} and \eqref{eq:IntC_arc}. Thus, the velocity vanishes at the gas-liquid interface and the Lagrange multiplier $c_0$ is the gradient of the surface-pressure within the incompressible surface-fluid. In the following, we will use this velocity field as the starting point of a perturbation expansion in the dimensionless deflection $\varepsilon = h(0)/a$ at the center of the gas-liquid interface.

We remark that the boundary condition \eqref{eq:BC_arc} is compatible with a constant Marangoni stress opposing the main flow in the limit of large Marangoni number when surface compressiblity is small. In this case the surfactant concentration remains virtually constant on the entire interface and even small gradients in surfactant concentration lead to large stresses opposing a compression of the surfactant layer. This is briefly explored in appendix \ref{sec:SurfaceIncompressibility}.

\subsection{Dimensionless formulation}\label{seq:dimlessForm}

Using the length-scale $a$ and the velocity scale $u_0=a\tau_\infty/\mu$ allows introducing dimensionless coordinates $(X,Y)=(x/a,y/a)$ and a dimensionless velocity $W(X,Y)=w(aX,aY)/u_0$ in $\Omega$. The integral boundary condition \eqref{eq:IntC_arc} then reads
\begin{equation}\label{eq:noDim_IntC_arc}
    \int_{\cal S} W(X,Y) dS = 0,
\end{equation}
and the no-slip condition on the planar surface $\overline{{\cal L}}$, condition \eqref{eq:BC_arc} on the gas-liquid interface ${\cal S}$, and the far-field condition \eqref{eq:farField} become
\begin{alignat}{2}
W(X,0) &= 0,	 																					&\quad&\text{on } \bar{\cal L}, \label{eq:noDim_noSlip}\\
\mathbf{n}\cdot \widetilde{\boldsymbol{\nabla}} W(X,Y) &= C,&&\text{on } {\cal S}, 				\label{eq:noDim_S}\\
\partial_Y W(X,Y) &\to 1,	 															&&\text{for } Y \to \infty,		\label{eq:noDim_Inf}
\end{alignat}
where $C=\mu c/\tau_\infty$ is the dimensionless Lagrange multiplier and $\widetilde{\boldsymbol{\nabla}}$ is the gradient in the dimensionless coordinates $(X,Y)$. Since $W$ solves the Laplace equation \eqref{eq:Laplace},
\begin{equation}\label{eq:noDim_Laplace}
\widetilde{\boldsymbol{\nabla}}\vphantom{\nabla}^2 W = 0,\quad \text{in }\Omega,
\end{equation}
shear-driven Stokes flow can be written as the imaginary part of an analytic function
\begin{equation}
f(Z) = V(X,Y) +i W(X,Y), \qquad Z=X+iY,
\end{equation}
by introducing the complex coordinate $Z=X+iY$. Denoting the complex derivative of $f$ as $f'$, and writing $\Re[\cdot]$ and $\Im[\cdot]$ for real and imaginary parts, the Cauchy-Riemann conditions, 
$\Re[f']=\partial_X V= \partial_Y W$ and $\Im[f']=-\partial_Y V=\partial_X W$, allow rewriting the boundary conditions \eqref{eq:noDim_noSlip}-\eqref{eq:noDim_Inf} as conditions for $f$ and its derivative.

\subsection{Domain perturbation}

We parameterize the deflection $y=h(x)$ of the gas liquid interface ${\cal S}$ by its maximal dimensionless deflection $\varepsilon = h(0)/a$ at its center. The radius of curvature of its circular arc then becomes $r=(a^2+h(0)^2)/(2h(0))=a(\varepsilon^{-1}+\varepsilon)/2$, counted as positive for $\varepsilon>0$, that is, when the deflection is into the upper half-plane. Up to second order in $\varepsilon$ the deflection then becomes
\begin{equation}\label{eq:deflection}
h(x) = \sigma(\varepsilon)\left( \sqrt{r^2-x^2} - \sqrt{r^2-a^2} \right)= \varepsilon a \left(1-\left(\frac{x}{a}\right)^2\right) + {\cal O}\left(\varepsilon^3\right),
\end{equation}
where $\sigma(\varepsilon)=\{1 \text{ for } \varepsilon>0; -1 \text{ for } \varepsilon<0; 0 \text{ for } \varepsilon=0 \}$ is the sign function. Thus, the dimensionless deflection $H(X)=h(aX)/a$ has the expansion
\begin{equation}
	H(X) = \varepsilon H_1(x/a) + {\cal O}\left(\varepsilon^3\right) = \varepsilon\left(1-X^2\right) + {\cal O}\left(\varepsilon^3\right).
\end{equation}

For small values of the dimensionless deflection $\varepsilon$, the velocity field $W$ can be found using a domain perturbation method \citep{Leal_2007}, by projecting the boundary conditions \eqref{eq:noDim_S} and \eqref{eq:noDim_IntC_arc} onto $\cal L$, the projection of ${\cal S}$ into the real axis (c.f. figure \ref{fig:sketch}(a)). For this, we use a regular perturbation expansion in $\varepsilon$ for $f$, $W$ and $C$,
\begin{align}
f(Z) &= Z + \varepsilon f_1(Z) + \varepsilon^2 f_2(Z) + \cdots \\
W(X,Y) &= Y + \varepsilon W_1(X,Y) + \varepsilon^2 W_2(X,Y) + \cdots \\
C &= 1 + \varepsilon C_1 + \varepsilon^2 C_2 + \cdots,
\end{align}
where we have used the fact that for a flat interface $f_0(Z)=Z$, $W_0(X,Y) = Y$ and $C_0=1$. The boundary condition \eqref{eq:noDim_S} projected onto ${\cal L}$ then becomes up to second order in $\varepsilon$
\begin{equation}\label{eq:epsBC_arc}
\begin{split}
\mathbf{n}\cdot \tilde{\boldsymbol{\nabla}} W \big|_{\cal S} &= \frac{-H'(X) \partial_X W(X,Y)+\partial_Y W(X,Y)}{\sqrt{1+(H'(X))^2}}\bigg|_{Y=H(X)} \\
& = 1	+ \varepsilon \partial_Y W_1(X,0)	\\
& \quad + \varepsilon^2 \left[ \partial_Y W_2(X,0) - \partial_X \big(H_1(X)\partial_X W_1(X,0)\big) - \tfrac{1}{2}\big(H_1'(X)\big)^2 \right] + \cdots \\
& = 1 + \varepsilon C_1 + \varepsilon^2 C_2 + \cdots
\end{split}
\end{equation}
where the Laplace equation \eqref{eq:noDim_Laplace} was used in the second equality. Similarly, for the projected integral boundary condition \eqref{eq:noDim_IntC_arc} we obtain
\begin{equation}\label{eq:epsIntBC_arc}
\begin{split}
0 &= \int_{-1}^1 W(X,H(X))\sqrt{1+\left(H'(X)\right)^2} dX 
\\
&= \int_{-1}^1 \left[ \varepsilon \left(W_1(X,0) + H_1(X) \right)
	+ \varepsilon^2 \left(W_2(X,0) + H_1(X)\partial_Y W_1(X,0) \right)
	\right] dX + \cdots.
\end{split}
\end{equation}
Note that the measure for the arc-length containing the square-root does not contribute to the expansion to order $\varepsilon^2$ in \eqref{eq:epsIntBC_arc}.

\subsection{Solution}\label{subseq:solution}

At each order in $\varepsilon$ we successively seek solutions for $f_i(Z)$ and $W_i(X,Y)$ in the whole upper half plane $\Omega_0=\{Z\,|\,\Im[Z]\geq 0\}$, obeying the boundary conditions \eqref{eq:noDim_noSlip} on the no-slip surface $\overline{\cal L}$, \eqref{eq:noDim_Inf} in the far field and the projections \eqref{eq:epsBC_arc} and \eqref{eq:epsIntBC_arc} on ${\cal L}$.  

\subsubsection{Order $\varepsilon^0$}
For the flat surface we already established
\begin{equation}
f_0(Z) = Z, \qquad C_0=1.
\end{equation}

\subsubsection{Order $\varepsilon^1$}
Using \eqref{eq:noDim_noSlip}, \eqref{eq:epsBC_arc} and \eqref{eq:noDim_Inf}, $W_1(X,Y)$ obeys the boundary conditions 
\begin{alignat}{2}
W_1(X,0) &= 0	&\quad&\text{on } \bar{\cal L},\\
\partial_Y W_1(X,0) &= C_1 &&\text{on } {\cal L},\\
\partial_Y W_1(X,Y) &\to 0	&&\text{for } Y \to \infty.
\end{alignat}
We recognize this as the boundary conditions for the flow above an infinite plane, driven by a constant shear rate $C_1$ along a strip of constant width on an otherwise no-slip surface. The well known solution is  \citep{Philip_1972}
\begin{equation}\label{eq:sol_f1}
f_1(Z) = C_1 \left(Z - \sqrt{Z^2-1} \right).
\end{equation}
Note that $f_1(Z)$ and hence $W_1(Z)$ vanish far from the surface. While the far-field boundary condition explicitly only forces the shear rate to vanish, this behavior is expected, as the momentum flux occurs between the surface of the groove (where momentum is introduced) and the surrounding no-slip surface. Inserting $W_1$ into \eqref{eq:epsIntBC_arc}, we obtain for the Lagrange multiplier
\begin{equation}
C_1 = \frac{8}{3\pi}.
\end{equation}

\subsubsection{Order $\varepsilon^2$}
The boundary conditions for $W_2$ inferred from \eqref{eq:noDim_noSlip}, \eqref{eq:epsBC_arc} and \eqref{eq:noDim_Inf} become
\begin{alignat}{2}
W_2(X,0) &= 0	&\quad&\text{on } \bar{\cal L},\\
\partial_Y W_2(X,0) &= \tfrac{1}{2}\left(H_1'(X)\right)^2 + \partial_X\left(H_1(X)\partial_X W_1(X,0) \right) + C_2 &\quad&\text{on }{\cal L},\\
\partial_Y W_2(X,Y) &\to 0	&\quad&\text{for } Y \to \infty.
\end{alignat}
It is instructive to rewrite these as conditions on the real and imaginary parts of the analytic function $f_2'$, using the fact that $\partial_X W_2 = 0$ on $\overline{\cal L}$, together with the Cauchy-Riemann conditions, $H_1(X)=1-X^2$, and \eqref{eq:sol_f1}
\begin{alignat}{2}
\Im[f'_2(X)] &= 0	&\quad&\text{on } \bar{\cal L},\label{eq:BCfPrime1}\\
\Re[f'_2(X)] &= 2X^2 + C_1 \frac{1-2X^2}{\sqrt{1-X^2}} + C_2 &\quad&\text{on }{\cal L},\label{eq:BCfPrime2}\\
\Re[f'_2(Z)] &\to 0	&\quad&\text{for } Z \to X+i\infty.\label{eq:BCfPrime3}
\end{alignat}
We recognize this as a mixed boundary value problem for $f'_2(Z)$ on the real line, which can be converted into a Riemann-Hilbert problem for which the methods of solution are well developed \citep{Gakhov_1966, Muskhelishvili_2008, Lawrentjew_1967}. In particular, using the Keldysh-Sedov formula (\cite{Gakhov_1966}, section 46.3; see also Appendix \ref{sec:RiemannHilbert}) we obtain
\begin{align}\label{eq:KeldyshSedov}
f'_2(Z) &= \frac{1}{i\pi}\frac{1}{\sqrt{Z^2-1}}\int_{-1}^{1} \frac{\sqrt{X^2-1}}{X-Z}\left(2X^2 + C_1 \frac{1-2X^2}{\sqrt{1-X^2}} + C_2\right) dX + \frac{A}{\sqrt{Z^2-1}} \\
\begin{split}\label{eq:sol_f2prime}
&= \frac{2 \left(\sqrt{Z^2-1}-Z\right) Z^2 + Z}{\sqrt{Z^2-1}}+C_1\frac{\left(2 Z^2-1\right) \log\left(\frac{Z+1}{Z-1}\right)-4Z}{\pi \sqrt{Z^2-1}} 
\\
&\qquad + C_2 \left(1-\frac{Z}{\sqrt{Z^2-1}}\right) + \frac{A}{\sqrt{Z^2-1}}.
\end{split}
\end{align}
The line integrals above can be performed by standard techniques \citep{England_2003, Gogolin_2014b, Muskhelishvili_2013}, and the term proportional to $A$ is a solution obeying homogeneous boundary conditions (i.e., $\Im[f'(X)]=0$ on $\overline{\cal L}$, $\Re[f'(X)]=0$ on ${\cal L}$, $f'(X+i\infty)=0$). The function $f_2$ can be obtained by taking the antiderivative of \eqref{eq:sol_f2prime}, 
\begin{equation}
\begin{split}\label{eq:sol_f2}
f_2(Z) &= 
\frac{2 \left(\sqrt{Z^2-1}-Z\right) Z^3+Z^2+1}{3 \sqrt{Z^2-1}} 
\\
& \quad + C_1\frac{\sqrt{Z^2-1} \left(Z\log\left(\frac{Z+1}{Z-1}\right) - 2\right)}{\pi }
+ C_2 \left(Z - \sqrt{Z^2-1} \right)
\end{split}
\end{equation}
where we have taken into account that, just as $W_1$, the velocity field $W_2$ vanishes in the far field, requiring $A=0$. The Lagrange multiplier $C_2$ is obtained from \eqref{eq:epsIntBC_arc} as
\begin{eqnarray}
C_2 = \frac{1}{2}\left(\left(\frac{8}{3\pi}\right)^2-1\right).
\end{eqnarray}
This concludes our determination of the velocity field as a perturbation series up to second order in the dimensionless deflection $\varepsilon$. As an aside, we note that the solution \eqref{eq:sol_f1} at order $\varepsilon^1$ also appears as the last part of the solution \eqref{eq:sol_f2} at order $\varepsilon^2$, corresponding to a constant shear rate $C_2$ instead of $C_1$ on ${\cal L}$.

\subsection{Branch cuts and choice of analytic functions}\label{sec:branchCuts}
\begin{figure}
	\centering
	\includegraphics[width=0.85\textwidth]{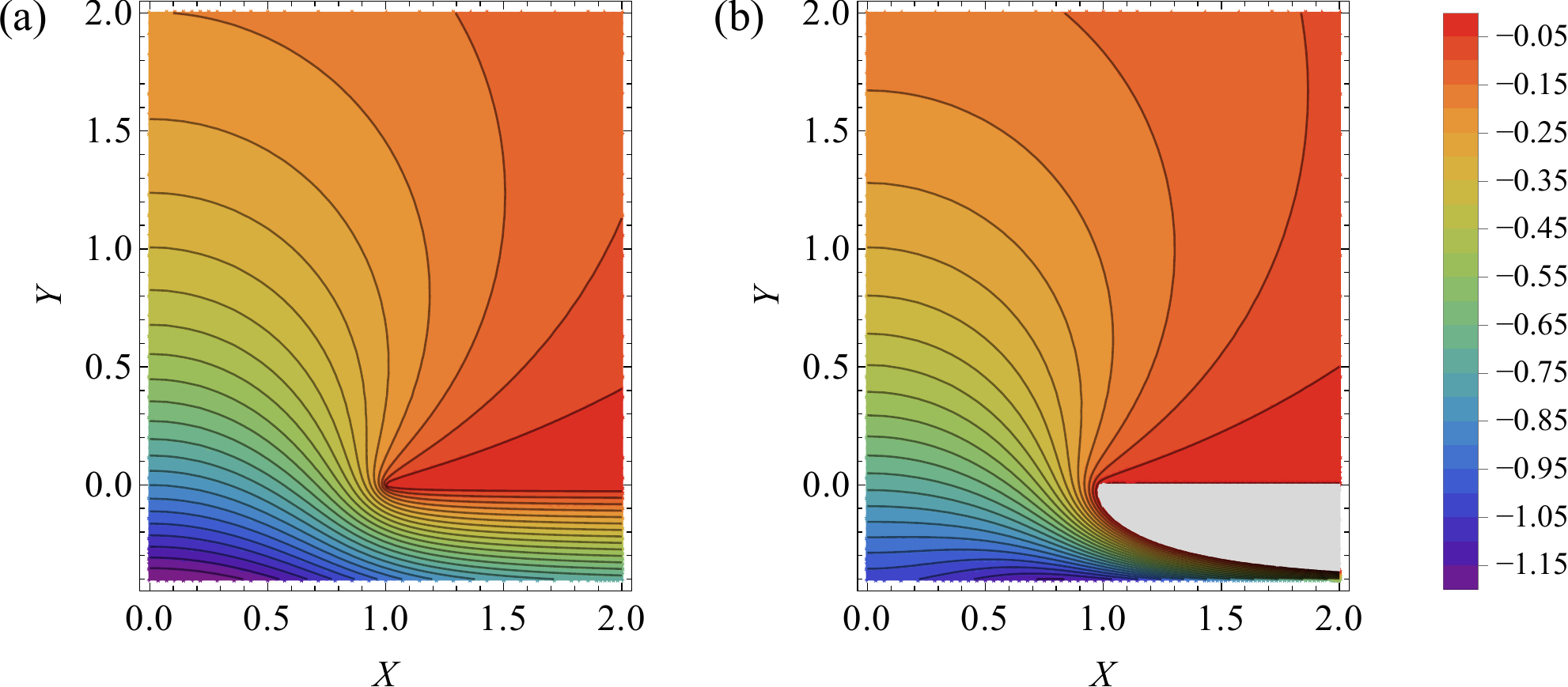}
	\caption{\label{fig:f1f2} Contour plot of (a) $\Im [f_1(Z)]$ and (b) $\Im [f_2(Z)]$. In the gray region $\Im [f_2(Z)]>0$; since here $X>1$ and $Y<0$, this region lies outside the considered liquid domain. The interval between contour lines is 0.05.}
\end{figure}

The principal branches of the analytic functions $\sqrt{Z}$ and $\log(Z)$ both have branch cuts on the negative real axis. By the nature of the solution method, some of the functions in our solution are discontinuous across the section $-1<X<1$ of the real axis. However, for $\varepsilon<0$ the functions sought after must be analytic across this line segment. We thus replace
\begin{align}
\sqrt{Z^2-1} & \to i\sqrt{1-Z^2} \\
\log\left(\frac{Z+1}{Z-1}\right) &\to \log\left(\frac{1+Z}{1-Z}\right)-i\pi,
\end{align}
which have branch cuts along the real axis \textit{except} on $-1<X<1$, agree with their original expressions in the first quadrant, and thus solve the Laplace equation in the region of interest while obeying the same boundary conditions when approaching the real line from above. We will assume these replacements to have been carried out in the expressions \eqref{eq:sol_f1} and \eqref{eq:sol_f2} for $f_1$ and $f_2$ without further mentioning them. In order to illustrate the resulting expressions, we have plotted their imaginary parts, i.e. their contribution to the velocity, in figure \ref{fig:f1f2}. In order to simplify the presentation we restrict the plot of $\Im [f_2(Z)]$ to negative values, since the corresponding region (shown in gray) is of no interest for the problem at hand. Above the real axis both functions have a very similar appearance, while marked differences are apparent for $Y<0$. This is an indication that our approximation may be better for positive deflection, $\varepsilon > 0$, than for negative deflection.  

\subsection{Finite element calculation}

For comparison with the analytical results, numerical calculations were performed using the commercial finite-element solver COMSOL Multiphysics (version 5.6, COMSOL AB, Stockholm, Sweden), employing its 'Coefficient form PDE' interface. The symmetry of the problem with respect to reflection at the $Y$-axis allows us to restrict the calculation to the region with $X\ge 0$ of the liquid domain of figure \ref{fig:sketch}(a). Specifically, the Laplace equation, \eqref{eq:noDim_Laplace}, was solved in a quadratic domain $0<X,Y<D$ of width and height $D = 25$  with a circular-arc section, corresponding to the deflected gas-liquid interface, added below (or removed above) the $X$-axis at $X<1$. On the circular arc the integral conservation equation \eqref{eq:noDim_IntC_arc} is prescribed as a constraint, while a Dirichlet condition, $W=0$, enforces the no-slip condition on the rest of the bottom surface. A constant shear rate $\partial_Y W = 1$ is applied on the top surface at $Y=D$, and a vanishing shear rate, $\partial_X W = 0$, is assumed on the left and right edges at $X=0$ and $D$, corresponding to a symmetry condition. The domain is discretized using quadratic Lagrange elements on a triangular mesh with cells of size $h_B=0.05$ away from the surface and $h_S=h_B/5$ on the circular arc, with a maximal element growth rate of 1.01. It was verified that in the range $-0.3 < \varepsilon < 0.4$ the velocity at the center of the interface at $(X,Y)=(0,\varepsilon)$ changes by less than 0.01\,\% when quadrupling the domain size $D$ and by less than 0.4\,\% when halving the element size $h_B$ and $h_S$, indicating that the results are independent of the grid, and the influence of the finite extent of the domain plays no significant role.

\section{Results and discussion}

\begin{figure}
	\centering
	\includegraphics[width=0.8\textwidth]{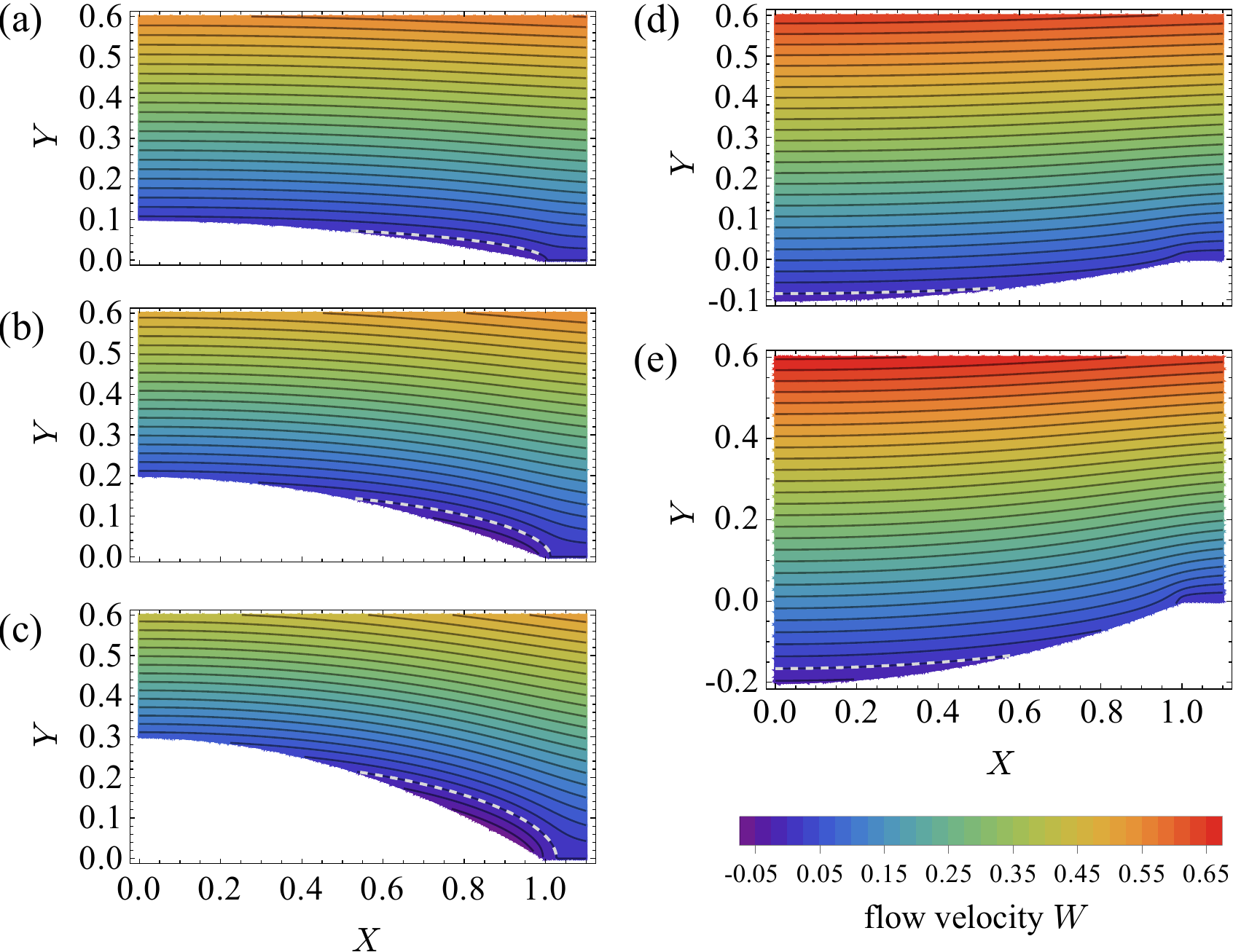}
	\caption{\label{fig:u2D} Contour plot of velocity $W(X,Y)$ for (a) $\varepsilon=0.1$, (b) $\varepsilon=0.2$, (c) $\varepsilon=0.3$, (d) $\varepsilon=-0.1$, (e) $\varepsilon=-0.2$. The interval between contour lines is 0.025 and the contour $W(X,Y)=0$ is shown as a dashed white line.}
\end{figure}

Velocity fields $W(X,Y)=\Im[f(X+iY)]$ are shown in figure \ref{fig:u2D} for $\varepsilon$ varying between -0.2 and 0.3 in steps of 0.1. Due to the symmetry of the problem only half of the cavity and its immediate vicinity is shown. As can be seen, for $\varepsilon >0$ the velocity at the center of the interface near $X=0$ is positive and becomes negative towards the edge of the cavity close to $X=1$. By contrast, for $\varepsilon<0$ a backflow is induced at the center of the cavity, while the flow velocity is positive towards the edge of the cavity. It is also apparent that the velocity on the interface is relatively small, and when leaving the interface into the fluid domain is quickly dwarfed by the increasing velocity due to the constant shear rate applied in the far field. As expected, compared to a pure Couette flow, the velocity profile in the vicinity of the groove attains slightly larger values for negative deflection and slightly smaller values for positive deflection. Due to the smallness of the interface velocity the situation is not much different from the case of a solid protrusion into the channel. Note that without incompressible surface fluid the dimensionless velocity at the center of a flat ($\varepsilon=0$) shear-free interface is 1 \citep{Philip_1972}, and thus the corresponding velocity field in the vicinity of the groove becomes markedly different.

\begin{figure}
	\centering
	\includegraphics[width=\textwidth]{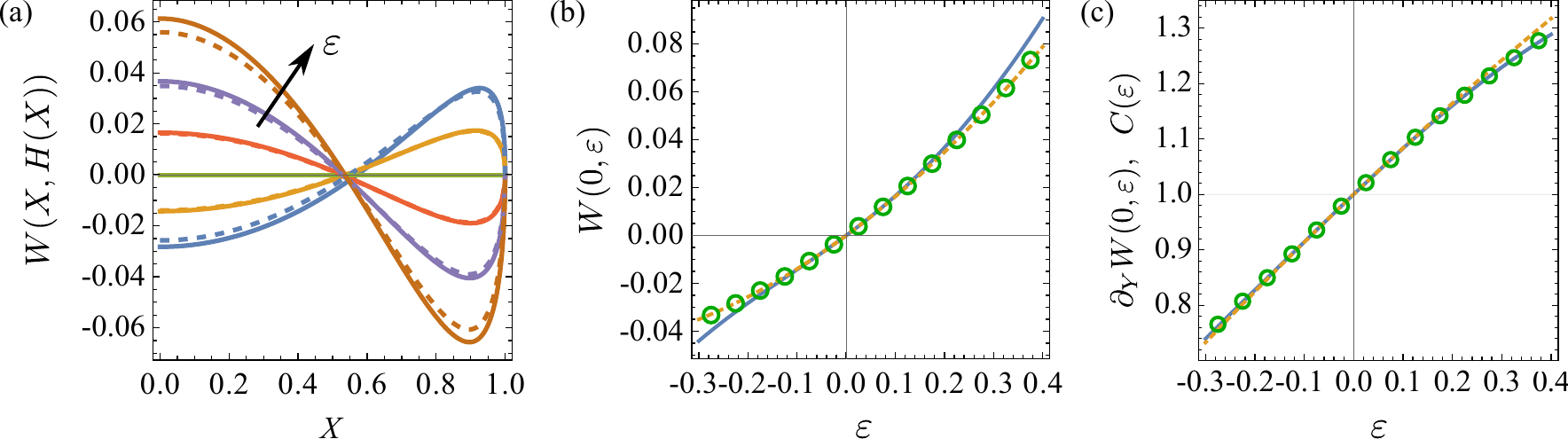}
	\caption{\label{fig:W_S_dW_S} (a) Velocity on interface, $W(X,\varepsilon H_1(X))$ (solid lines) and the corresponding numerical results (dashed lines) for $\varepsilon=-0.2$, -0.1, 0, 0.1 0.2, 0.3. The arrow indicates increasing values of $\varepsilon$. (b) Velocity $W(0,\varepsilon)$ on the interface at $X=0$ (blue solid line), projected velocity $\tilde W_S(0)$ (equation \eqref{eq:eps_W_projected}, yellow dashed line) and corresponding numerical values (green circles). (c) Shear rate, $\partial_Y W(0,\varepsilon)$, on the interface at $X=0$ (blue solid line), $C(\varepsilon)$ (yellow dashed line) and corresponding numerical values (green circles).}
\end{figure}

The velocity on the interface is more clearly shown in figure \ref{fig:W_S_dW_S}(a), where it is plotted for the same values of $\varepsilon$ as in figure \ref{fig:u2D}. Here it becomes particularly transparent that with an incompressible surface fluid the velocity only reaches a few percent of the values that would be reached on a shear-free interface. At the same time, the corresponding velocities from the numerical calculations using the finite element discretisation are shown as dashed lines. It is evident that the analytical solution agrees well with the numerical calculation in the chosen range of $\varepsilon$, with the largest deviations occurring close to the center of the cavity. The fact that backflow occurs on the interface, with flow in opposite directions close to the center of the groove and close to its edges, is evidently a prerequisite for the mass conservation of the incompressible surface fluid. Its direction of transport at the center reflects the fact that a deflection of the interface above or below the reference plane leads to a respectively increased or decreased shear rate on the interface compared to the planar surface, see figure \ref{fig:W_S_dW_S}(c). 

To further assess the quality of the analytical solution, the velocity close to the center of the cavity is plotted as a function of $\varepsilon$ in figure \ref{fig:W_S_dW_S}(b) (blue line) together with corresponding values from the numerical calculations (green circles). Note that the quality of the approximation is not symmetric in $\varepsilon$ and deviations become lager more quickly for negative deflections than for positive ones. This expected behavior was already alluded to above in the discussion of the functions $f_1(Z)$ and $f_2(Z)$ in section \ref{sec:branchCuts}.

Another estimate for the range of validity of the expansion can be obtained by investigating the expansion of $W_S(X) = W(X,\varepsilon H_1(X))$ to second order in $\varepsilon$ as
\begin{equation}\label{eq:eps_W_projected}
\begin{split}
\tilde W_S(X) &= \varepsilon \left(1-X^2 - C_1 \sqrt{1-X^2} \right) 
+ \varepsilon^2 \Bigg( 
-\frac{1}{3}\sqrt{1-X^2}\left(1 + 2X^2 + 3C_2\right) \\
&\hspace{1cm}+C_1\left[1-X^2-\frac{\sqrt{1-X^2}}{\pi}\left\{2+X\log\left(\frac{1-X}{1+X}\right)\right\}\right]
\Bigg).
\end{split}
\end{equation}
Since similar expansions are performed during the projection of boundary conditions in \eqref{eq:epsBC_arc} and \eqref{eq:epsIntBC_arc}, the difference between $\tilde W_S(0)$ and $W(0,\varepsilon)$ should indicate when the expansion becomes inadequate. Remarkably, $\tilde W_S(0)$, shown as a yellow dashed line in figure \ref{fig:W_S_dW_S}(b), closely traces the numerical values obtained and is thus a faithful indicator of the range of validity of our expansion.

Equation \eqref{eq:eps_W_projected} thus is an excellent approximation for determining the extremal velocity encountered at the midpoint of the channel in the interval $-0.2\lesssim \varepsilon \lesssim 0.3$. It can also be used to assess the extremum occuring towards the edges of the groove. When considering \eqref{eq:eps_W_projected} to first order in $\varepsilon$ only, the location of this extremum lies at $X=\sqrt{1-C_1^2/4}\approx 0.905$, while for the full equation \eqref{eq:eps_W_projected} the location $X \approx 0.9 \pm 0.02$ of this extremum is only weakly dependent on $\varepsilon$ in the considered interval of $\varepsilon$ and thus $\tilde{W}_S(0.9)$ remains within 1\% of the value obtained from \eqref{eq:eps_W_projected} at the location of the extremum. Similarly, the position where the interfacial velocity changes sign varies little with $\varepsilon$ and according to \eqref{eq:eps_W_projected} lies between $X=0.54$ and 0.51 in the considered interval of $\varepsilon$, tightly straddling the position $X=\sqrt{1-C_1^2}\approx 0.529$ of the zero of equation \eqref{eq:eps_W_projected} to first order in $\varepsilon$.

Another value of interest is the shear rate on the interface. In our expansion this can be obtained directly from $\partial_Y W(0,\varepsilon)=\Re[f'(i\varepsilon)]$. At the same time, from equation \eqref{eq:noDim_S}, the shear rate at the interface is encoded in the Lagrange multiplyer $C(\varepsilon)$. Again, any deviation between the two is an indication for the quality of our approximation. The curves for both these quantities are shown as the respective blue solid and yellow dashed lines in figure \ref{fig:W_S_dW_S}(c), together with the numerically obtained values as green circles. Note that in the numerical calculations the shear is constant on the entire interface. As can be seen, the shear rate at the center of the interface is less sensitive to the approximation than the velocity. However, for the shear rate the discrepancies are expected to become largest towards the edges of the groove, and this is indeed reflected in our solution (not shown).

\section{Conclusion and Outlook}

We have presented an analytical solution for shear-driven liquid flow along a single bounded gas-filled groove, embedded in an otherwise planar surface, when the gas-liquid interface protrudes slightly above or below the planar surface and is covered by an incompressible surface fluid. The flow velocities displayed in figures \ref{fig:u2D} and \ref{fig:W_S_dW_S}(a) show that backflow occurs close to the edges of the groove when the interface protrudes above the planar surface and close to its center when it deflects below the plane. Evidently, such regions with backflow are mandatory since on a bounded groove as much surface fluid is transported along the flow direction as against it.

Qualitatively the same behavior of the interface flow predicted for an incompressible surface fluid was observed experimentally at a single gas-filled groove in the presence of large concentrations of surfactant by \cite{Song_2018}. We take this as evidence for the adequacy of our model for describing such situations. Nevertheless, there are some discrepancies, as experimentally the backflow on the interface is not always as prominent as in our prediction. This may be due to interfacial concentrations not being large enough for the surfactant film to become fully incompressible or due to surfactant dissolving in the liquid. Moreover,  experiments were performed using pressure-driven Poiseuille flow in a relatively shallow channel instead of pure shear flow, affecting the details of the boundary conditions. While a more complex model for surfactants taking into account adsorption/desorption kinetics and an equation of state or an effective surface viscosity at large surfactant concentrations could be incorporated in a more complete model, it would be difficult to capture such details in an analytical description. Despite the mentioned shortcomings, we hope that our analytic model will be a valuable reference for describing flow with large interfacial concentrations of surfactants.

Flow over a superhydrophobic surface in Cassie state is often characterized by reporting an apparent slip length \citep{Rothstein_2010, Lee_2016}. It should therefore be of interest to extend the present study from a single groove to a parallel array of gas-filled grooves in order to investigate the impact of large surfactant concentrations on the observable slip length. Moreover, as mentioned in the introduction, thermal Marangoni flow along a grooved surface can also be approximately described by a constant shear stress along the grooves \citep{Baier_2010}. The similarity to the flow over an incompressible surface fluid promises some synergy between the investigations of both situations. 

More generally, flow over arbitrary gas- or liquid-filled patches (e.g. circular holes) covered by an incompressible surface fluid, with interfaces protruding above or below a flat surface is of interest for various designs of superhydrophobic surfaces. In this respect, an extension to a slightly compressible surface fluid, possibly exhibiting surface-viscosity, may be of interest for a more complete picture. Such configurations may also be relevant in other situations such as flow over vesicles attached to a solid wall \citep{Woodhouse_2012, Honerkamp_2013}.

On a higher level of abstraction, we hypothesize that the results reported in this paper may hint at a quite general class of fluid dynamic phenomena: that the flow along a surfactant-covered liquid surface is very sensitive to the surface deformation. In our case, the surface flow is suppressed on a flat surface, while a flow emerges on a deformed surface. It is conceivable that the flow field itself deforms a liquid surface, or that the surface deformation is controlled by an external parameter (such as pressure), which in turn would influence the flow pattern on the surface.

\begin{acknowledgments}
\end{acknowledgments}

Declaration of Interests. The authors report no conflict of interest.

\appendix
\section{Surface incompressibility}\label{sec:SurfaceIncompressibility}

In section \ref{sec:model} the interface was modelled as containing an incompressible surface fluid. In this appendix we briefly discuss the compatibility of this simple model with momentum conservation and transport of a surfactant species at the interface in the limit of large Marangoni number.

The steady-state interfacial stress balance in the Boussinesq-Scriven model reads \citep{Edwards_1991, Slattery_2007, Manikantan_2020}
\begin{equation}\label{eq:Boussinesq-Scriven}
\mathbf{n} \cdot \left(p^{(+)} - p^{(-)}\right) + \gamma(\boldsymbol{\nabla}_s\cdot\mathbf{n})\mathbf{n} = \mathbf{n} \cdot \left(\boldsymbol{\tau}^{(+)} - \boldsymbol{\tau}^{(-)}\right) -\boldsymbol{\nabla}_s \mathbf{\Pi} + \boldsymbol{\nabla}_s\cdot\boldsymbol{\tau}_s
\end{equation}
where $\mathbf{I}_s = \mathbf{I}-\mathbf{n}\mathbf{n}$ is the interface projection operator and $\boldsymbol{\nabla}_s = \mathbf{I}_s\cdot\boldsymbol{\nabla}$ the interface gradient.  $p^{(+)}$, $p^{(-)}$, $\boldsymbol{\tau}^{(+)}$ and $\boldsymbol{\tau}^{(-)}$ are the pressures and viscous stress tensors in the fluid on the side of the interface the normal vector $\mathbf{n}$ points to and away from, respectively, with $\boldsymbol{\tau}^{(\cdot)}=\mu^{(\cdot)}(\boldsymbol{\nabla}\mathbf{u}^{(\cdot)}+(\boldsymbol{\nabla}\mathbf{u}^{(\cdot)})^T)$. On the interface it is assumed that the velocities of both fluid phases are identical, $\mathbf{u}=\mathbf{u}^{(+)}=\mathbf{u}^{(-)}$. $\Pi(\Gamma) = \gamma_0 - \gamma(\Gamma)$ is the surface pressure with $\gamma(\Gamma)$ being the interfacial tension of an interface with surfactant concentration $\Gamma$ and $\gamma_0$ the interfacial tension of the clean interface. The interfacial reheology is captured in the Boussinesq-Scriven stress
\begin{equation}\label{eq:surfaceStress}
\boldsymbol{\tau}_s = [(\kappa_s-\mu_s) \boldsymbol{\nabla}_s \cdot \mathbf{u}]\mathbf{I}_s + \mu_s [\boldsymbol{\nabla}_s\mathbf{u}\cdot\mathbf{I}_s+\mathbf{I}_s\cdot(\boldsymbol{\nabla}_s\mathbf{u})^T],
\end{equation}
where $\mu_s$ and $\kappa_s$ is the surface shear and the surface dilatational viscosity, respectively. 

In our model with a gas-liquid interface, we assume that the viscous stress $\boldsymbol{\tau}^{(-)}$ on the gas side is negligible compared to the viscous stress $\boldsymbol{\tau}^{(+)}$ on the liquid side. For small Boussinesq numbers $\Bq_\mu=\mu_s/(\mu a)$ and $\Bq_\kappa=\kappa_s/(\mu a)$, the intrinsic surface stresses can be neglected compared to the stresses exerted by the adjacent fluid. In our case the $z$-component of \eqref{eq:Boussinesq-Scriven} along the grove then reads
\begin{equation}\label{eq:momentum_z}
\mu \mathbf{n}\cdot\boldsymbol{\nabla} w 
= \partial_z \mathbf{\Pi} 
= \frac{\partial \Pi(\Gamma)}{\partial \Gamma} \partial_z \Gamma 
= \Gamma_0 \frac{\partial \Pi(\Gamma)}{\partial \Gamma} \partial_z \delta \widetilde{\Gamma},
\end{equation}
where in the last step we have assumed that the surfactant concentration has the form $\Gamma=\Gamma_0(1+\delta\widetilde{\Gamma})$ with $\delta \widetilde{\Gamma} \ll 1$. It is useful to introduce the Marangoni modulus \citep{Manikantan_2020}
\begin{equation}
E_0(\Gamma) = \Gamma \frac{\partial \Pi}{\partial \Gamma},
\end{equation}
as a measure of the interfacial elasticity or the amount of work needed for compressing an interface with surfactants. As in section \ref{seq:dimlessForm} we use the length scale $a$ for introducing dimensionless coordinates, here writing $\tilde{Z}=z/a$ for the $z$-coordinate, and the velocity scale $u_0=a\tau_\infty/\mu$. The non-dimensional form of equation \eqref{eq:momentum_z} is approximately
\begin{equation}\label{eq:stressBalance}
\mathbf{n}\cdot\widetilde{\boldsymbol{\nabla}} W = \frac{\Gamma_0}{\mu u_0} \frac{\partial \Pi(\Gamma)}{\partial \Gamma} \partial_{\tilde{Z}} \delta\widetilde{\Gamma} \approx \Ma \, \partial_{\tilde{Z}} \delta\widetilde{\Gamma},
\end{equation}
where we have introduced the Marangoni number 
\begin{equation}
\Ma = \frac{E_0(\Gamma_0)}{\mu u_0} = \frac{E_0(\Gamma_0)}{a \tau_\infty}.
\end{equation}
Consequently, for a constant shear stress along the groove, the surfactant gradient becomes arbitrarily small for large $\Ma$, with $\delta\widetilde{\Gamma}$ remaining small for not too large groove lengths. Since the left-hand side of \eqref{eq:stressBalance} is constant in our model, this equation corresponds to equation \eqref{eq:BC_arc} (or \eqref{eq:noDim_S}).

The surfactant flux at the interface in $z$-direction along the grove is governed by convection and diffusion,
\begin{equation}\label{eq:longitudinalTransport}
\begin{split}
N_{\Gamma,z} &= w \Gamma - D \partial_z \Gamma 
= u_0 \Gamma_0 \left(W (1+\delta\widetilde{\Gamma}) - \Pe^{-1} \partial_{\tilde{Z}} \delta\widetilde{\Gamma}\right) \\
&\approx u_0 \Gamma_0 \left(W\left(1+\Ma^{-1}\mathbf{n}\cdot\widetilde{\boldsymbol{\nabla}} W \widetilde{Z}\right) - (\Pe\,\Ma)^{-1} \mathbf{n}\cdot\widetilde{\boldsymbol{\nabla}} W\right) \approx w\Gamma_0,
\end{split}
\end{equation}
where we have substituted $\delta\widetilde{\Gamma}$ according to \eqref{eq:stressBalance} and in the last step taken the limit of large Marangoni number, assuming that $\tilde{Z}=z/a$ is not too large and $Pe=au_0/D$ is not too small. Additionally, in our approximation the transverse velocity on the groove vanishes, which is consistent with a negligible gradient in the surfactant concentration in this direction. Thus, for large $\Ma$ the surfactant concentration can be considered as constant, $\Gamma_0$, on the interface, with $\Pi$ taking the role of the pressure in the momentum equation for the incompressible surface fluid. Integrating \eqref{eq:longitudinalTransport} across the width of the interface then leads to equation \eqref{eq:IntC_arc} of our model. Finally, the pressure difference between the phases is governed by the mean curvature $H$ of the interface, $2H=-\boldsymbol{\nabla}_s\cdot\mathbf{n}$, such that $p^{(+)} - p^{(-)} = 2 \gamma H$. As for large $\Ma$ the surface tension $\gamma$ stays nearly constant on the interface, this is consistent with the circular arc cross section of the interface assumed in our model.

Typical values of the shear rates employed in experiments are $\dot{\gamma}=\tau_\infty/\mu=0.1 \ldots 1 \text{ s}^{-1}$, for water with $\mu \simeq 1$~mPa~s, and a typical length scale is $a = 0.1 \ldots 1$~mm. With the Marangoni modulus scaling as $E_0 \simeq k_B T \Gamma_0$ at a surfactant concentration $\Gamma_0 \simeq 0.01 \ldots 1 \text{ nm}^{-2}$, typical values for the Marangoni number lie in the range $\Ma \simeq k_B T \Gamma_0 / (a \tau_\infty) \simeq 4\cdot (10^1 \ldots 10^5)$. With surface diffision coefficients in the range $D \simeq (0.1\ldots 1)\cdot 10^{-9}$~m~s$^{-2}$ the corresponding Peclet numbers are in the range $\Pe = au_0/D\simeq a^2 \dot{\gamma}/D\simeq 1 \ldots 10^4$. It is thus expected that in many experimental scenarios the assumption of an incompressible surfactant phase is well justified.

\section{Outline of a derivation of the Keldysh-Sedov formula} \label{sec:RiemannHilbert}

In section \ref{subseq:solution} a holomorphic function was obtained in the upper half-plane, obeying certain boundary conditions on the real line, by using the Keldysh-Sedov formula. Here we give a brief sketch how to obtain this expression based on the behaviour of holomorphic functions at cuts on the real line. Standard references for these techniques are \cite{Muskhelishvili_2008} or \cite{Gakhov_1966} while \cite{England_2003} gives an excellent brief introduction.

Consider a function $\Phi(Z)$ that is holomorphic in the complex plane with the exception of possible cuts on intervals located on the real line. For such a function it is useful to define the limit when approaching a point $X$ on the real line from above or below
\begin{equation}
    \Phi^\pm(X) = \lim_{\varepsilon\to 0^{+}} \Phi(X \pm i\varepsilon).
\end{equation}
When $\Phi(Z)$ has purely real values in an interval on the real line, the Schwartz reflection principle, $\overline{\Phi(Z)}=\Phi(\bar{Z})$, where the overbar denotes complex conjugation, leads to $\overline{\Phi^+(X)}=\Phi^-(X)$. As an example, the function
\begin{equation}
    R(Z) = \sqrt{Z^2-1}
\end{equation}
has a cut on the interval $[-1,1]$ of the real line and obeys
\begin{equation} \label{eq:jumpR}
\begin{aligned}
R^+(X) &= -R^-(X)	            \quad\text{for } X^2 \leq 1,\\
R^+(X) &=\phantom{-}  R^-(X)    \quad\text{for } X^2 > 1.
\end{aligned}
\end{equation}

According to the Sokhotski–Plemelj theorem \citep[section~1.4.1]{Gogolin_2014b} 
\begin{equation}
    \lim_{\varepsilon\to 0^{+}}\int_{-\infty}^\infty \frac{f(X)dX}{X \pm i\varepsilon} 
    = \mp i\pi f(0) + \mathcal{P}\int_{-\infty}^\infty \frac{f(X)dX}{X},
\end{equation}
where $\mathcal{P}$ denotes the principal value of the integral. Hence the function
\begin{equation}\label{eq:CauchyLine}
    F(Z) = \frac{1}{2\pi i} \int_{-\infty}^\infty \frac{f(X)dX}{X - Z}
\end{equation}
is another example that obeys the jump condition
\begin{equation}\label{eq:Plemelj}
    F^+(X) - F^-(X) = f(X).
\end{equation}

The conditions \eqref{eq:jumpR} and \eqref{eq:Plemelj} can be used to obtain a holomorphic function $g(Z)$ in the upper half-plane that obeys boundary conditions as in \eqref{eq:BCfPrime1} and \eqref{eq:BCfPrime2}
\begin{equation}
\begin{aligned}
\Re[g^+(X)] &= \tfrac{1}{2}[ g^+(X) + g^-(X) ] = \eta(X) &\quad\text{for } X^2 \leq 1,\\
\Im[g^+(X)] &= \tfrac{1}{2}[ g^+(X) - g^-(X) ] = 0    &\quad\text{for } X^2 > 1.
\end{aligned}
\end{equation}
Introducing $\psi(Z) = R(Z)g(Z)$ and using \eqref{eq:jumpR}, these can be converted to
\begin{equation}
    \psi^+(X) - \psi^-(X) = \left\{ 
    \begin{aligned}
    &2R^+(X)\eta(X) &\quad\text{for } X^2 \leq 1, \\
    &0           &\quad\text{for } X^2 > 1,
    \end{aligned}
    \right.
\end{equation}
which has the form of \eqref{eq:Plemelj}. Finally, using \eqref{eq:CauchyLine}, we obtain
\begin{equation}
    \psi(Z) = \sqrt{Z^2-1} \, g(Z) = \frac{1}{2\pi i} \int_{-1}^1 \frac{\sqrt{X^2-1}\,\eta(X)dX}{X - Z}.
\end{equation}
Apart from the solution of the homogeneous problem, this is the Keldysh-Sedov formula of \eqref{eq:KeldyshSedov}. Further generalisations and details on the conditions for applicability can be found in the standard references listed above.


\bibliographystyle{jfm_abbr1stName} 
\bibliography{references}

\end{document}